\documentclass[a4paper,twoside,12pt]{article}
\usepackage[a4paper,margin=2cm]{geometry}
\usepackage[utf8]{inputenc}
\usepackage[T1]{fontenc}
\usepackage{lmodern}
\usepackage[pdftex]{hyperref}

\title{HAR in Smart Homes}
\author{Damien Bouchabou, Christophe Lohr, Ioannis Kanellos, Sao Mai Nguyen}
\date{November 19, 2021}

\begin{document}
\maketitle

\section*{Definition}
Human Activity Recognition (HAR) consists in monitoring and analyzing the
behavior of one or more persons in order to deduce their activity.  In a
smart home context, the HAR consists in monitoring daily activities of the
residents, based on a network of IoT devices.  Owing to this monitoring, a
smart home can offer personalized home assistance services to improve
quality of life, autonomy and health of their residents, especially for
elderly and dependent people.

\paragraph{Keywords:} survey; human activity recognition; deep learning;
smart home; ambient assisting living;taxonomies; challenges; opportunities;
HAR

\section{Introduction}
Human Activity Recognition (HAR) from sensors consists in using a network of
sensors and connected devices to track a person's activity.  This produces
data in the form of a time series of state changes or parameter values.  The
wide range of sensors --contact detectors, RFID, accelerometers, motion
sensors, noise sensors, radar, etc.-- can be placed directly on a person,
on objects or in the environment.  The sensor-based solutions are generally
divided into three categories: Wearable, Sensors on Objects and Ambient
Sensors \cite{ref1}.

Tracking at home people's activities can pose serious privacy issues.  While
camera installation can be part of various security services, residents are
generally reluctant to leave cameras and monitoring systems turned on when
they are home.  Sensor-based systems have dominated the applications of
daily activities recognition in smart homes insofar as they are globally
less intrusive \cite{ref2,ref3}.  Thanks to the development of the Internet
of Things and the multiplication of cheap and powerful smart devices, smart
homes based on ambient sensors have become a viable technical solution, in
order to offer various services.  Beyond the hardware, the latter also need
algorithms able to exploit such a potential.

Many  surveys \cite{ref2,ref3,ref4,ref5,ref6,ref7} involve this research
question.  But they rarely tackled smart homes problems.  Yet, HAR in smart
homes is a crucial and challenging problem because human activity is complex
and variable from a resident to another.  Every resident has different
lifestyle, habits or abilities.  The wide range of daily activities, as well
as the variability and the ﬂexibility in how they can be performed, requires
an approach both scalable and adaptive.  Algorithms for HAR in smart homes
and the challenges for the ambient sensors studied in HAR can be classiﬁed
as pertaining to a problem of pattern classiﬁcation, of temporal data
analysis or of data variability.  The proposed taxonomy is described below.

\section{Pattern Classification}
Algorithms for HAR in smart homes are ﬁrst pattern recognition algorithms. 
Based on characteristics and criteria, a method identiﬁes patterns in order
to assign them a category.  The methods found in the literature can be
divided into two broad categories: Knowledge-Driven Approaches (KDA) and
Data-Driven Approaches (DDA)

\subsection{Knowledge-Driven Approaches} In KDA, an activity model is built
through the incorporation of rich prior knowledge gleaned from the
application domain.  This is done by using knowledge engineering and
knowledge management techniques.  KDA are motivated by real-world
observations that involve activities of daily living and lists of objects
required for performing such activities.  In real life situations, even if
the activity is performed in different ways, the number and the type of
involved objects do not vary signiﬁcantly.  For example, the activity ``to
brush teeth'' contains actions involving a toothbrush, a toothpaste, a water
tap, cup, and a towel.  On the other hand, as humans have different
lifestyles, habits and abilities, they can perform an activity in different
ways.

KDA also uses the observation that most activities, in particular, routine
activities of daily living and working, take place in certain circumstances
of time and location.  For example, ``brushing teeth'' is generally
undertaken twice a day in a bathroom, in the morning and before going to bed
and involves a minima the use of toothpaste and toothbrush.  These implicit
relations, which build up a local universe on the basis of singular actions,
temporal and spatial data and involved objects, provide a diversity of hints
and foster heuristics for inferring activities.

KDA for HAR are actually ontology-based approaches.  They are commonly used
as ontological activity models that do not depend on algorithmic choices. 
They have been thoroughly used to construct reliable activity models.  One
can ﬁnd in a comprehensive overview of such approaches.  Ontologies can be
used to represent objects in activity spaces exploiting, noticeably, the
semantic relations between objects and activities, like in \cite{ref8}. 
Such approaches aim at automatically detecting possible activities related
to an object.

\subsection{Data-Driven Approaches}
DDA include both supervised and unsupervised machine learning methods, which
primarily use probabilistic and statistical reasoning.  The DDA strength is
the probabilistic modeling capacity.  These models are capable of handling
noisy, uncertain, and incomplete sensor data.  They can capture domain
heuristics, e.g., some activities are more likely than others.  They do not
require to set up a predeﬁned domain knowledge.  However, DDA require much
data and, in the case of supervised learning, clean and correctly labeled
data.

Several classiﬁcation algorithms were evaluated.  Boundaries classiﬁers,
such as, Decision Trees \cite{ref8,ref9}, Conditional Random Fields, or
Support Vector Machines have been used.  Probabilistic classiﬁers, such as
the Naive Bayes classiﬁer, also showed good performance in learning and
classifying oﬄine activities when a large amount of training data is
available.

Most of these methods use hand crafted features, similar to KDA that use
prior knowledge.  Generating features by hand is time consuming, and avoids
adaptability.  Another DDA axis, the Deep Learning (DL) approaches (CNN
\cite{ref10}, FCN \cite{ref11} , Auto-encoders \cite{ref12}, LSTM
\cite{ref13}, etc.), are now used to overcome this limitation of features
extraction and also perform the classiﬁcation task \cite{ref14}.

\section{Temporal Data}
Smart home sensors produce a data stream of events that are ordered in time. 
This stream is unfortunately not sampled in a regular way like other time
series problems.  Indeed, it is an event stream because sensors are
activated or change state when the resident interacts or performs an action. 
A sensor activation corresponds more or less to a resident's action.  There
may be a few seconds or a several minutes between events.  It is this
sequence of events that translates into a sequence of actions and therefore
an activity.  It is worth noting the challenges of dealing with the temporal
complexity of human activity data in real use cases.

Activities can be more or less complex.  A simple activity is an activity
that consists of a single action or movement, such as walking, running,
turning on the light or opening a drawer.  A complex activity is an activity
that involves a sequence of actions, potentially involving different
interactions with objects, equipment or other people, as, for example,
cooking.

Clearly, if monitoring the activities of daily living performed by a single
resident is already a complex task, the complexity increases drastically
when we have to deal with several residents.  The same activities become
harder to recognize.  On the one hand, in a group, a resident may interact
to perform common activities.  In this case, the activation of the sensors
reﬂects the same activity for each resident in the group.  On the other
hand, everyone can perform different activities simultaneously.  This
produces a simultaneous activation of the sensors for different activities. 
These activations are then merged and combined in the activity sequences. 
An activity performed by some resident may become a noise for the activities
of another.

Long Short Term Memory (LSTM) algorithms show excellent performance on the
classiﬁcation of irregular time series in the context of a single resident
and simple activities.  However, human activity is much more complex than
this.  Long-term dependencies often occur in activities of daily living.  To
tackle this non-Markovian time series, a context can be introduced to help
the understanding of the observed activations. \cite{ref15} used language models to
both estimate the duration of this dependency and encode this context. 
Challenges related to the recognition of concurrent, interleaved or idle
activities offer more diﬃculties.  Currently, work on HAR in smart homes
does not take into account these types of activities.  Moreover, people,
generally, do not live alone in a house.  This is why even more complex
challenges are introduced, including the recognition of activity in homes
with multiple residents.  These challenges, which address multi-class
classiﬁcation problems, are still unsolved.

\section{Data Variability}
The complexity of real human activities is not the sole problem.  Indeed,
the application of human activity recognition in smart homes for real-use
cases faces moreover issues causing a sound discrepancy between training and
test data.  Some of these issues are inherent to smart homes: the temporal
drift of the data and the variability of settings.

\subsection{Temporal drift}
To accommodate this drift, algorithms for HAR in smart homes should
incorporate life-long learning to continuously learn and adapt to changes in
human activities from new data.  Recent works in life-long learning
incorporating deep learning, as reviewed in \cite{ref16}, could help tackle this
issue of temporal drift.  In particular, one can imagine that an interactive
system can from time to time request labeled data to users, to continue to
learn and adapt.  Such algorithms have been developed under the names of
interactive reinforcement learning or active imitation learning in robotics. 
For instance, in \cite{ref17}, the system is allowed to learn micro and compound
actions, while minimizing the number of requests for labeled data by
choosing when and what information to ask, and even to whom to ask for help. 
Such principles could inspire a smart home system to continue to adapt its
model, while minimizing user intervention and optimizing his intervention,
by pointing out the missing key information.

\subsection{Variability of settings} Beside these long-term evolutions, the
data from one house to another are also very different.  Thus, the model
learned in one house is hardly applicable in another because of the change
in house conﬁguration, sensors equipment and families' compositions and
habits.  As a matter of fact, the location, the number and the sensor type
of smart homes can inﬂuence activity recognition performances of a system. 
Smart homes can be equipped in different ways and may have different
architectures in terms of sensors, room conﬁguration, appliance, etc
\cite{ref18,ref19}.  Some can have a lot of sensors, multiple bathrooms, or
bedrooms and contain multiple appliances, while others can be smaller, such
as a single apartment, where sensors can be fewer and have more overlaps and
noisy sequences.  Due to this difference in house conﬁgurations, a model,
optimized for a particular smart home, could perform poorly in another.  Of
course, this issue can be faced by collecting a new dataset for each new
household, in order to train the models anew; however, this is costly.  A
less costly solution for data augmentation would be to collect the data in a
simulated appartement.

Another solution is to adapt the models learned in a household to another. 
Transfer learning methods have recently been developed to allow pre-trained
deep learning models to be used with different data distributions,. 
Transfer learning using deep learning has been successfully applied to time
series classiﬁcation, as reviewed in \cite{ref19}.  For activity
recognition, \cite{ref20} reviewed the different types of knowledge that
could be transferred in traditional machine learning.  These methods can be
updated with deep learning algorithms and by taking advantage of current
researches in transfer learning for deep learning.  Furthermore, adaptation
to new settings have recently been improved by the development of
meta-learning algorithms.  Their goal is to train a model on a variety of
learning tasks, so it can solve new learning tasks using only a small number
of training samples.  This ﬁeld has seen recent breakthroughs, as reviewed
in \cite{ref21}, which has never been applied yet to HAR \cite{ref22}.  Yet,
the peculiar variability of data of HAR in smart homes can only hope of some
beneﬁt from such algorithms.

\section{Conclusion}
HAR in smart homes have demonstrated interesting advances owed, mainly, to
the development of recent Deep Learning algorithms for end-to-end
classiﬁcation such as convolutional neural networks.  It also beneﬁts from
recent algorithms for sequence learning such as long-short term memory. 
However, as with video processing, sequence learning still needs to be
improved to be able, both, to deal with the vanishing gradient problem and
to take into account the context of the sensor readings.  The temporal
dimension is incidentally a particularity of ambient sensor systems, as the
data is a sparse and irregular time series.  The irregular sampling in time
has also been tackled with adapted windowing methods for data segmentation. 
In addition to the time windows used in other HAR ﬁelds, sensor event
windows are commonly used as well.  The sparsity of the data of ambient
sensors do not allow machine learning algorithms to take advantage of the
redundancy of data over time, as in the case of videos where successive
video frames are mostly similar.  Moreover, whereas HAR in videos the
context of the human action can be seen in the images by the detection of
the environment or the objects of attention, the sparsity of the HAR in
ambient sensors result in a high reliance on the past information to infer
the context information.

While HAR in ambient sensors has to face the problems of complex activities
such as sequences of activities, concurrent activities or multi-occupant
activities, or even data drift, it has also to tackle speciﬁc unsolved
problems such as the variability of data.  Indeed, the data collected by
sensors are even more sensitive to the house conﬁguration, the choice of
sensors and their localization.

\bigskip

This entry is adapted from \href{http://doi.org/10.3390/s21186037}{10.3390/s21186037}.

\bigskip
Retrieved from \url{https://encyclopedia.pub/17108}


\section*{References}
\begin{thebibliography}{-------}

\bibitem{ref1}
Li, X.; Zhang, Y.; Marsic, I.; Sarcevic, A.; Burd, R.S.
\newblock Deep learning for rfid-based activity recognition. 
\newblock In Proceedings of the 14th ACM Conference on Embedded Network Sensor Systems, Stanford, CA, USA, 14--16 November 2016; pp. 164--175.

\bibitem{ref2}
Gomes, L.; Sousa, F.; Vale, Z.
\newblock An intelligent smart plug with shared knowledge capabilities.
\newblock Sensors 2018, 18, 3961.      

\bibitem{ref3}
Gochoo, M.; Tan, T.H.; Liu, S.H.; Jean, F.R.; Alnajjar, F.S.; Huang, S.C.
\newblock Unobtrusive activity recognition of elderly people living alone using anonymous binary sensors and DCNN.
\newblock IEEE J. Biomed. Health Inform. 2018, 23, 693--702.

\bibitem{ref4}
Yan, S.; Lin, K.J.; Zheng, X.; Zhang, W.
\newblock Using latent knowledge to improve real-time activity recognition for smart IoT.
\newblock  IEEE Trans. Knowl. Data Eng. 2020, 32, 574--587.  

\bibitem{ref5}
Perkowitz, M.; Philipose, M.; Fishkin, K.; Patterson, D.J. 
\newblock Mining models of human activities from the web. 
\newblock In Proceedings of the 13th International Conference on World Wide Web, New York, NY, USA, 17--20 May 2004; pp. 573--582. 

\bibitem{ref6}
Chen, L.; Nugent, C.D.; Mulvenna, M.; Finlay, D.; Hong, X.; Poland, M. 
\newblock A logical framework for behaviour reasoning and assistance in a smart home. 
\newblock Int. J. Assist. Robot. Mechatron. 2008, 9, 20--34. 

\bibitem{ref7}
Chen, L.; Nugent, C.D. 
\newblock Human Activity Recognition and Behaviour Analysis.
\newblock Springer: Berlin/Heidelberg, Germany, 2019. 

\bibitem{ref8}
Fleury, A.; Vacher, M.; Noury, N. 
\newblock SVM-based multimodal classification of activities of daily living in health smart homes: Sensors, algorithms, and first experimental results.
\newblock IEEE Trans. Inf. Technol. Biomed. 2009, 14, 274--283.

\bibitem{ref9}
Cook, D.J. 
\newblock Learning setting-generalized activity models for smart spaces. 
\newblock IEEE Intell. Syst. 2010, 2010, 1.

\bibitem{ref10}
Tan-Hsu Tan; Munkhjargal Gochoo; Shih-Chia Huang; Yi-Hung Liu; Shing-Hong Liu; Yung-Fa Huang.
\newblock  Multi-Resident Activity Recognition In A Smart Home Using RGB Activity Image and DCNN.
\newblock IEEE Sensors Journal 2018, PP, 1-1, 10.1109/jsen.2018.2866806.

\bibitem{ref11}
Damien Bouchabou; Sao Mai Nguyen; Christophe Lohr; Benoit LeDuc; Ioannis Kanellos.
\newblock  Fully Convolutional Network Bootstrapped by Word Encoding and Embedding for Activity Recognition in Smart Homes. 
\newblock Communications in Computer and Information Science 2021, 1, 111-125, 10.1007/978-981-16-0575-8\_9.

\bibitem{ref12}
Aiguo Wang; Guilin Chen; Cuijuan Shang; Miaofei Zhang; Li Liu.
\newblock Human Activity Recognition in a Smart Home Environment with Stacked Denoising Autoencoders.
\newblock Lecture Notes in Computer Science 2016, 1, 29-40, 10.1007/978-3-319-47121-1\_3.

\bibitem{ref13}
Daniele Liciotti; Michele Bernardini; Luca Romeo; Emanuele Frontoni.
\newblock A sequential deep learning application for recognising human activities in smart homes.
\newblock Neurocomputing 2019, 396, 501-513, 10.1016/j.neucom.2018.10.104.

\bibitem{ref14}
Jindong Wang; Yiqiang Chen; Shuji Hao; Xiaohui Peng; Lisha Hu.
\newblock Deep learning for sensor-based activity recognition: A survey. 
\newblock Pattern Recognition Letters 2019, 119, 3-11, 10.1016/j.patrec.2018.02.010.

\bibitem{ref15}
Damien Bouchabou; Sao Mai Nguyen; Christophe Lohr; Benoit LeDuc; Ioannis Kanellos.
\newblock Using Language Model to Bootstrap Human Activity Recognition Ambient Sensors Based in Smart Homes.
\newblock Electronics 2021, 10, 2498, 10.3390/electronics10202498.

\bibitem{ref16}
Sebastian Thrun; Lorien Pratt.
\newblock Learning to Learn: Introduction and Overview. 
\newblock Learning to Learn 1998, 1, 3-17, 10.1007/978-1-4615-5529-2\_1.

\bibitem{ref17}
Thrun, S.; Pratt, L. 
\newblock Learning to Learn; Springer US: Boston, MA, USA, 1998. 

\bibitem{ref18}
Parisi, G.I.; Kemker, R.; Part, J.L.; Kanan, C.; Wermter, S. 
\newblock Continual lifelong learning with neural networks: A review.
\newblock  Neural Netw. 2019, 113, 54--71.  

\bibitem{ref19}
Duminy, N.; Nguyen, S.M.; Zhu, J.; Duhaut, D.; Kerdreux, J. 
\newblock Intrinsically Motivated Open-Ended Multi-Task Learning Using Transfer Learning to Discover Task Hierarchy.
\newblock Appl. Sci. 2021, 11, 975.  

\bibitem{ref20}
Weiss, K.; Khoshgoftaar, T.M.; Wang, D. 
\newblock A survey of transfer learning. 
\newblock Big Data 2016, 3, 9. 

\bibitem{ref21}
Fawaz, H.I.; Forestier, G.; Weber, J.; Idoumghar, L.; Muller, P.A. 
\newblock Transfer learning for time series classification. 
\newblock In Proceedings of the 2018 IEEE International Conference on Big Data (Big Data), Seattle, WA, USA, 10--13 December 2018; pp. 1367--1376. 

\bibitem{ref22}
Cook, D.; Feuz, K.; Krishnan, N. 
\newblock Transfer Learning for Activity Recognition: A Survey. 
\newblock Knowl. Inf. Syst. 2013, 36, 537--556.   

\bibitem{ref23}
Daniele Liciotti; Michele Bernardini; Luca Romeo; Emanuele Frontoni.
\newblock A sequential deep learning application for recognising human activities in smart homes. 
\newblock Neurocomputing 2019, 396, 501-513, 10.1016/j.neucom.2018.10.104.

\bibitem{ref24}
Damien Bouchabou; Sao Mai Nguyen; Christophe Lohr; Benoit LeDuc; Ioannis Kanellos.
\newblock Using Language Model to Bootstrap Human Activity Recognition Ambient Sensors Based in Smart Homes.
\newblock Electronics 2021, 10, 2498, 10.3390/electronics10202498.

\end{thebibliography}
\end{document}